\documentclass[aps,showpacs,preprint,epsf,epsfig, nofootinbib,titlepage]{revtex4-1}
\usepackage{graphicx}% Include figure files
\usepackage{dcolumn}% Align table columns on decimal point
\usepackage{bm}% bold math
\usepackage{hyperref}% add hypertext capabilities
\usepackage{epsfig,graphicx,color,appendix}
\usepackage{multirow}
\usepackage{array}
\usepackage{tabularx}
\usepackage{capt-of}
\usepackage{amsmath}
\usepackage{caption}
\usepackage[countmax]{subfloat}

\begin{document}

\def\|{\vert}
\def\ol{\bar}
\def\ov{\bar}
\def\be{\begin{eqnarray}}
\def\en{\end{eqnarray}}
\def\non{\nonumber}
\def\t{\times}
\def\lb{\Lambda}
\def\om{\Omega}
\def\omc{\Omega_c}
\def\oc{\Omega_c^0}
\def\xic{\Xi_c}
\def\xicd{\Xi_c^{'}}
\def\x{\Xi}
\def\si{\Sigma}
\def\k{{K_1}}
\def\k1d{\underline{K}_{1}}
\def\kb{\bar{K}_{1}}
\def\kbd{\underline{\bar K}_{1}}
\def\ra{\rangle}
\def\A{{\cal A}}
\def\B{{\cal B}}
\def\L{{\cal L}}
\def\ben{\begin{eqnarray}}
\def\en{\end{eqnarray}}
\def\non{\nonumber}
\def\la{\langle}
\def\ra{\rangle}
\def\t{\times}
\def\pp{{\prime\prime}}
\def\nc{N_c^{\rm eff}}
\def\vp{\varepsilon}
\def\hep{\hat{\varepsilon}}
\def\up{\uparrow}
\def\dw{\downarrow}
\def\vma{{_{V-A}}}
\def\vpa{{_{V+A}}}
\def\smp{{_{S-P}}}
\def\spp{{_{S+P}}}
\def\J{{J/\psi}}
\def\ov{\bar}
\def\Lqcd{{\Lambda_{\rm QCD}}}
\long\def\symbolfootnote[#1]#2{\begingroup%
\def\thefootnote{\fnsymbol{footnote}}\footnote[#1]{#2}\endgroup}
\def\lsim{ {\ \lower-1.2pt\vbox{\hbox{\rlap{$<$}\lower5pt\vbox{\hbox{$\sim$}
}}}\ } }
\def\gsim{ {\ \lower-1.2pt\vbox{\hbox{\rlap{$>$}\lower5pt\vbox{\hbox{$\sim$}
}}}\ } }

\font\el=cmbx10 scaled \magstep2{\obeylines\hfill \today}

\title{ \Large Axial-Vector Emitting \\
Weak Nonleptonic Deacys of  \boldmath $\Omega_c^0$ Baryon \\
\mbox{}}

\author{\bf Rohit Dhir$^{\dagger}$  and C. S. Kim$^{\ddagger}$ }
\email{$^{\dagger}$dhir.rohit@gmail.com, $^{\ddagger}${cskim@yonsei.ac.kr} }

\affiliation{\sl  Department of Physics and IPAP, Yonsei University, Seoul 120-749, Korea
\\
\\
\\
\\
\\
}
 \null
\null\null\null

\begin{abstract}
\noindent The axial-vector emitting weak hadronic decays of $\Omega_c^0$ baryon are investigated.
 After employing the factorization and the pole model framework to predict their branching ratios, we derive the symmetry breaking effects on axial-vector-meson-baryon couplings
 and effects of flavor dependence on baryon-baryon weak transition amplitudes and, consequently, on their branching ratios. We found that the W-exchange process contributions dominate $p$-wave meson emitting decays of $\omc^0$ baryon.
\end{abstract}

\maketitle
\thispagestyle{empty}

\setcounter{page}{1}

\section{INTRODUCTION}

 The production of heavy baryons have always  posed experimental challenges and hence, have generated much interest in their studies \cite{1,2,3,4}. Many interesting observations by CDF, D0, SELEX, FOCUS, Belle, BABAR, CMS, LHCb \textit{etc.} \cite{5,6,7,8,9,10,11,12,13,14} in context of mass spectrum, lifetimes and decay rates have been made in resent years. Most recently, LHCb and CDF collaborations \cite{15,16,17,18,19,20,21,22} have announced more precise measurement of masses and lifetimes of ($\x_c^0,~\x_c^+,~\lb_b^{(*)},~\x_b^-,~\x_b^0,~\Omega_b^-$) baryons. Also, LHCb has now identified two new strange-beauty baryonic resonances, denoted by $\x^{'-}_b$ and $\x^{*-}_b$ \cite{23}, though many of doubly and triply heavy states are yet to be confirmed. In two body nonleptonic decay sector, first observation of ($\Omega_b^- \to \oc \pi^-$) decay process and measurement of CP-asymmetries for $\lb_b \to p \pi^-$ and $ \lb_b \to p K^-$ are reported by CDF collaboration \cite{17,24}. On the other hand, LHCb has reported first observation of  $\lb_b \to \lb_c^+ D_{(s)}^-$ and $\lb_b \to J/\psi p \pi^-$ decays and the measurement of the difference in CP-asymmetries between $\lb_b \to J/\psi p \pi^-$ and $\lb_b \to J/\psi p K^-$ and many other decays involving $b-$baryons \cite{25,26,27,28,29}.  However, little progress has been made in observing decays of heavy charm meson. All these recent measurements have attracted, much needed, attention to heavy baryonic sector.

 On the theoretical side, various attempts had been made to investigate weak decays of heavy baryons \cite{30,31,32,33,34,35,36,37,38,39,40,41,42,43,44,45,46,47,48,49,50,51,52,53,54,55,56,57,58}. A number of methods, mainly, current algebra (CA) approach, factorization scheme, pole model technique,non-relativistic quark model (NRQM), heavy quark effective theory (HQET), framework based on next-to-leading order QCD improved Hamiltonian  \textit{etc.}, have been employed. Recent experimental developments have prompted more theoretical efforts in $b$-baryon decays \cite{58,59,60,61,62,63,64}. In all these works, the focus has so far been on \textit{s}-wave meson emitting decays of heavy baryons including $\oc$ decays. Being heavy, charm and bottom baryons can also emit \textit{p}-wave mesons. In past, the \textit{p}-wave emitting decays of charm and bottom baryons have been studied using factorization and pole model approach \cite{65,66,67,68,69,70}. However, \textit{p}-wave emitting decays of $\oc$ baryon remain untouched. The fact, that $\oc$ baryon is the heaviest and only doubly strange particle in charmed baryon sextet that is stable against strong and electromagnetic interactions, makes it an interesting candidate for the present analysis. Moreover, study of \textit{s}-wave emitting decays of  $\oc$ baryon reveals that nonfactorizable W-exchange terms dominate as compared to factorizable contributions \cite{37}. This makes study of $\oc$ decays even more important to understand the mechanism underlying W-exchange processes.

 In our previous work \cite{70}, we have studied the scalar meson emitting decays of bottom baryons employing the pole model. We have shown that such decays can acquire significant pole (W-exchange) contributions to make their branching ratios comparable to \textit{s}-wave meson emitting decays. In the present work, we analyze the axial-vector meson emitting exclusive nonleptonic decays of $\oc$ baryon. We have already seen that for $\oc$ decays the factorization contribution are small in comparison to the pole contributions in case of \textit{s}-wave meson emitting decays. Thus, the factorizable contributions to $p$-wave meson emitting decays of $\oc$ baryon are also expected to be suppressed. Therefore, we study weak nonleptonic decays of $\oc$ emitting axial-vector mesons in the factorization and the pole model approach. We obtain the factorization contributions using the non-relativistic quark model (NRQM) \cite{33} and heavy quark effective theory (HQET) \cite{47} based form factors. We employ the effects of symmetry breaking (SB) on strong couplings that may decide crucial pole diagram contributions \cite{50,71}. We use the traditional non-relativistic approach \cite{72} to evaluate weak matrix element to obtain flavor independent pole amplitude contributions at ground state $\frac{1}{2}^+$ intermediate baryon pole terms. Adding factorizable and pole contributions we predict branching ratios (BRs) of $\oc$ decays. Later, we employ the possible flavor dependence via variation of spacial baryon wave function overlap in weak decay amplitude. We find that BRs of all the decay modes are significantly enhanced on inclusion of flavor dependent effects.

 The article is organized as follows: In sec. II , we give a general framework including spectroscopy of axial-vector mesons, decay kinematics and effective Hamiltonian. Sec. III deals with weak decay amplitudes both pole terms and factorization terms, weak transitions and axial-vector meson-baryon couplings. Numerical results are given in sec IV. We summarize our findings in last section.

\section{GENERAL FRAMEWORK}

\subsection{Spectroscopy of Axial-Vector Mesons}

The axial-vector meson spectroscopy has extensively been studied in literatures \cite{73,74,75,76,77}. Here, we list the important facts. Spectroscopically, there are two types of axial-vector mesons: ${}^{3} P_{1} $($J^{PC} =1^{++} $) and ${}^{1} P_{1} $ $(J^{PC} =1^{+-} )$. ${}^{3} P_{1} $ and ${}^{1} P_{1} $  states can either mix within themselves or can mix with one another. Experimentally observed non-strange and uncharmed axial-vector mesons exhibit first kind of mixing and can be identified as follows:
\begin{description}
\item[${}^{3} P_{1} $] meson 16-plet includes isovector $a_{1}(1.230) $~\footnote{ Here the quantities in brackets indicate their respective masses (in GeV).} and four isoscalars, namely, $f_{1} (1.285)$,
$f_{1}(1.420)/f'_{1}(1.512)$ and $\chi _{c1} (3.511)$. The following mixing scheme has been used in isoscalar ($1^{++} $) mesons:
\ben f_{1} (1.285) &=& \frac{1}{\sqrt{2} } (u\overline{u}+d\overline{d})\cos \phi_{A} +(s\overline{s})\sin \phi_{A}, \non
  \\ f'_{1} (1.512)\,&=&\frac{1}{\sqrt{2} } (u\overline{u}+d\overline{d})\sin \phi_{A} -(s\overline{s})\cos \phi _{A}.
\en
\item[${}^{1} P_{1} $] meson multiplet consists isovector $b_{1} (1.229)$ and three isoscalars $ h_{1} (1.170)$, $ h'_{1} (1.380)$ and $h_{c1} (3.526)$, where spin and parities of $h_{c1} (3.526)$ and $ h'_{1} (1.380)$ states are yet to be confirmed, experimentally. These isoscalar ($1^{+-} $) mesons can mix in following manner:
\ben
 h_{1} (1.170)&=&\frac{1}{\sqrt{2} } (u\overline{u}+d\overline{d})\cos \phi {}_{A'} +(s\overline{s})\sin \phi {}_{A'}, \non \\  h'_{1} (1.380)&=&\frac{1}{\sqrt{2} } (u\overline{u}+d\overline{d})\sin \phi _{A'} -(s\overline{s})\cos \phi _{A'}.
\en
\end{description}
The mixing angles are given by relation: $\phi _{A(A')} =  \theta ({\rm ideal}) - \theta _{A(A')} ({\rm physical})$.
The experimental observations predominantly favor the ideal mixing for these states $i.e.$, $\phi _{A}  =\phi _{A'}  = 0^{\circ }$.

The hidden-flavor diagonal states $a_{1}(1.230) $  and $b_{1} (1.229)$ cannot mix owing to C- and G-parity considerations. However, there are no such restrictions for the states involving strange partners namely,  $K_{1A} $ and $K_{1A'} $ of $A\, (1^{++} )$ and $A'(1^{+-} )$ mesons, respectively. They mix in the following convention to generate the physical states :
\ben
 K_{1} (1.270)&=& K_{1A}  \sin \theta _{K_1}  + K_{1A'}  \cos \theta _{K_1} ,\non
  \\ \underline{K}_{1} (1.400)&=& K_{1A}  \cos \theta _{K_1}  - K_{1A'}  \sin \theta _{K_1} .
\en

  Several phenomenological analyses based on the experimental information obtained twofold ambiguous solutions for $\theta _{K_1} $ \textit{i.e.} $\pm \, 37^{\circ} $ and $\pm \, 58^{\circ} $ \cite{73,74,75,76,77}. We wish to point out that the experimental measurement of the ratio of $K_1 \gamma$ production in $B$ decays and the study of charm meson decays to $K_{1} (1.270) \pi / K_{1} (1.400) \pi $ favor negative angle solutions. Very recently \cite{75}, it has been shown that choice of mixing angle $\theta_{K_1}$ is intimately related to choice of angle for $f-f^{'}$ and $h-h^{'}$ mixing schemes. The mixing angle $\theta_{K_1} \sim 35^{\circ}$ is favored over $\sim 55^{\circ}$ for near ideal mixing for $f-f^{'}$ and $h-h^{'}$. Therefore, we use $\theta_{K_1}=-37^{\circ}$ for our calculation; however, we also give results on $-58^{\circ}$ for comparison.

\subsection{Kinematics}

The matrix element for the baryon decay process \textit{e.g.} $B_{i} (\frac{1}{2}^{+},~ p_i )\to B_{f} (\frac{1}{2}^{+},~ p_f )+A_{k} (1^{+},~q )$ can be expressed as
\ben \label{4} \la B_{f} (p_{f} )A_{k} (q)\vert H_{W} \vert B_{i} (p_{i} )\ra = i\bar{u}_{B_{f} } (p_{f}
)\varepsilon ^{*\mu } (A_{1} \gamma _{\mu } \gamma _{5} +A_{2} p_{f\mu }
\gamma _{5} +B_{1} \gamma _{\mu } +B_{2} p_{f\mu } )u_{B_{i} } (p_{i} ),
\en
where $u_{B} $ are Dirac spinors for baryonic states $B_i$ and $B_f$. $\varepsilon ^{\mu } $ is the polarization vector of the axial-vector
meson state $A_{k} $. $A_i$'s and $B_i$'s represent parity conserving (PC) and  parity violating (PV) amplitudes, respectively.
The decay width for the above process is given by
\ben
{\Gamma =\frac{q_{\mu } }{8\pi } \frac{E_{f} +m_{f} }{m_{i} } \Big[
2(|S|^{2} +|P_{2} |^{2} )+\frac{E_{A}^{2} }{m_{A}^{2} } (|S+D|^{2} +|P_{1}
|^{2} )\Big ] },
\en
where $m_{i} $ and $m_{f} $ are the masses of the initial and final state baryons, and $q_{\mu } =(p_{i} -p_{f} )_{\mu } $ is  the four-momentum of axial-vector meson
\ben|q_{\mu }| = \frac{1}{2m_i}\sqrt{[m_{i}^{2} -(m_{f} -m_{A} )^{2} ][m_{i}^{2} -(m_{f} +m_{A} )^{2} ]} ,\non \en
  where $m_{A} $ is the mass of emitted \textit{p}-wave meson \cite{34,41}. The decay emplitude of the final state  is now an admixture of \textit{S}, \textit{P} and \textit{D} wave angular momentum states with
\ben
S=-A_{1},~~P_{1} =-\frac{q_{\mu } }{E_{A} } \left(\frac{m_{i} +m_{f} }{E_{f} +m_{f} }
B_{1} +m_{i} B_{2} \right),
\non \en
\ben
P_{2} =\frac{q_{\mu } }{E_{f} +m_{f} } B_{1} ,~~D=-\frac{q_{\mu }^{2} }{E_{A} (E_{f} +m_{f} )} \left(A_{1} -m_{i} A_{2}
\right), \non
\en
where $E_A$ and $E_{f} $ are the energies of the axial-vector meson and the daughter baryon, respectively.
Furthermore, there are two
independent \textit{P}-wave amplitudes: one corresponds to the singlet
spin combination of the parent and daughter baryon and the other corresponds to the
triplet. The interference between  \textit{S }and \textit{D} wave amplitudes and  \textit{P}-wave amplitudes results in asymmetries for the
daughter state with respect to the spin of the parent state. The corresponding asymmetry parameter is
\ben
\alpha =\frac{4m_{A}^{2}~\mbox{Re}[S*P_{2}]+2E_{A}^{2}~\mbox{Re}[(S+D)*P_{1}]}{2m_{A}^{2}
(|S|^{2} +|P_{2} |^{2} )+E_{A}^{2} (|S+D|^{2} +|P_{1} |^{2} )}.\en
Thus, to determine the decay rate and asymmetry parameters we require to estimate amplitudes, $A$ and $B$.

\subsection{Hamiltonian}

The QCD modified current $\otimes $ current effective weak Hamiltonian consisting Cabibbo-favored ($\Delta C = \Delta S = -1$), Cabibbo-suppressed ($\Delta C = -1,  \Delta S = 0$) and Cabibo-doubly-suppressed ($\Delta C = - \Delta S = -1$) modes is given by
\ben
H_{W}^{eff} = \frac{G_{F}}{ \sqrt{2}} \big\{ V_{ud} V_{cs}^{*} &\big[& c_{1} (\bar{u}d)(\bar{c}s)+c_{2} (\bar{s}d)(\bar{u}c)\big]_{(\Delta C = \Delta S = -1)} + \non \\ V_{ud} V_{cd}^{*} &\big[& c_{1} \{(\bar{s}c)(\bar{u}s)-(\bar{d}c)(\bar{u}d)\}  + c_{2}\{ (\bar{u}c)(\bar{s}s)-(\bar{u}c)(\bar{d}d)\}\big]_{(\Delta C = -1,~  \Delta S = 0)} - \non \\ V_{us} V_{cd}^{*} &\big[& c_{1} (\bar{d}c)(\bar{u}s)+c_{2} (\bar{u}c)(\bar{d}s)\big]_{(\Delta C =-\Delta S= -1)}\big\} ,
\en
where $V_{ij}$  and $(\bar{q}_{i} q_{j} )\equiv \bar{q}_{i} \gamma _{\mu } (1-\gamma _{5} )q_{j} $ denote the Cabibbo-Kobayashi-Maskawa
(CKM) matrix elements and the weak \textit{V-A }current, respectively. We use the QCD coefficients  $c_{1} (\mu )=1.2$, $c_{2} (\mu )=-0.51$ at $\mu \approx m_{c}^{2} $ \cite{35}. Furthermore, nonfactorizable effects may modify $c_1$ and $c_2$, thereby indicating that these may be treated
as free parameters. The discrepancy between theory and experiment is greatly improved in the large $N_c$ limit. Interestingly, the charm conserving decays of $\oc$ are also possible but they are kinematically forbidden in present analysis.

\section{Decay Amplitudes}
The hadronic matrix element $\langle B_{f} A_{k}\vert H_{W} \vert B_{i} \rangle$ for the $B_i \to B_f +A_k$ process may be expressed as
\ben \langle B_{f} A_{k}\vert H_{W} \vert B_{i} \rangle \equiv \mathcal{A}_{Pole} + \mathcal{A}_{Fac.} ,  \en
where $\mathcal{A}_{Pole}$ and $\mathcal{A}_{Fac.}$ represent pole (W-exchange) and factorization amplitudes, respectively.
 The pole diagram contributions involving the W-exchange process are evaluated using the pole model framework \cite{35}.   One may consider factorization as a correction to the pole model which includes the calculation of possible pole diagrams via $s-$, $u-$ and $t-$channels, where $t-$channel virtually implicate tree-level diagrams $i.e.$ factorizable processes. The contribution of these terms can be summed up in terms of PC and PV amplitudes.

\subsection{Pole Amplitudes}
 The first term, $\mathcal{A}_{Pole}$, involves the evaluation of the relevant matrix element
\ben
\la B_f \vert H \vert B_i \ra = \bar{u}_{B_{i}} (B +\gamma_{5} A )u_{B_{j}}
\en
between two $\frac{1}{2}^+$ baryon states. \textit{A} and \textit{B} are $s-$wave  and $p-$wave decay amplitudes, receptively. $A$ and $B$ include the contributions of $s-$ and $u-$ channels for positive-parity intermediate baryon $(J^{P} =\frac{1}{2}^+ )$ poles; henceforth, given by $A^{pole}$ and $B^{pole}$ as follows:
\begin{equation}
A^{pole} =\mathop{-\Sigma }\limits_{n} \left[\frac{g_{B_{f} B_{n} A_{k} } a_{ni} }{m_{i} -m_{n} } +\frac{a_{fn} g_{B_{n} B_{i} A_{k} } }{m_{f} -m_{n} } \right],
\end{equation}
\begin{equation}
B^{pole} =\mathop{\Sigma }\limits_{n} \left[\frac{g_{B_{f} B_{n} A_{k} } b_{ni} }{m_{i} +m_{n} } +\frac{b_{fn} g_{B_{n} B_{i} A_{k} } }{m_{f} +m_{n} } \right],
\end{equation}
where $g_{ijk} $ are the strong axial-vector meson-baryon coupling constants; $a_{ij} $ and $b_{ij}$ are weak baryon-baryon matrix elements defined as
\begin{equation}
\la B_{i} \vert H_{W} \vert B_{j} \ra = \bar{u}_{B_{i} } (a_{ij} +\gamma _{5} b_{ij} )u_{B_{j} } .
\end{equation}
It is well known that the PV matrix element $b_{ij}$ vanishes for the hyperons owing to $\langle B_{f} A_{k}\vert H_{W}^{PV} \vert B_{i} \rangle =0$  in the SU(3) limit. This also implies for non-leptonic charm meson decays that $b_{ij} \ll a_{ij} $ suppressing \textit{s-}wave contributions for $\frac{1}{2}^+-$ poles. These contributions are further suppressed by presence of sum of the baryon masses in the denominator. Thus, only PC terms survive  for non-leptonic decays of charm baryons. It may be noted that the negative-parity intermediate baryon $(J^{P} = \frac{1}{2}^{-})$ may also contribute to these processes and may turn out to be important. However, evaluation of such terms require knowledge of the axial-vector meson strong coupling constants for $(\frac{1}{2}^{-})$ states. Unfortunately, no such theoretical or experimental information is available in literature. Moreover, \textit{in the leading non-relativistic approximation}, one can ignore $J^P=\frac{1}{2}^-,~ \frac{3}{2}^-....$ and higher resonances as they would require at least one power of momentum in $H_W$ in order to connect them with the relevant ground state in the overlap integral. That means one has consider terms of the order $v/c$. In the same manner, to connect radial excitations with the corresponding ground state, one would need terms of
order $(v/c)^2$; otherwise the overlap integral would be zero due to orthogonality of the wave functions \cite{51}. Therefore, we have restricted our calculation to ground state $\frac{1}{2}^+$ intermediate baryon pole terms to estimate the pole contributions to the axial-vector meson emitting decays of charm baryons.

\subsection{Factorizable Amplitudes}

Likewise meson decays, the reduced matrix element (\ref{4}) can be factorized to obtain decay amplitudes (ignoring the scale factors) in the following form:
\begin{equation}
<A_{k} (q)|A_{\mu } |0>\, <B_{f} (p_{f} )|V^{\mu } +A^{\mu } |B_{i} (p_{i} )>,
\end{equation}
where
\begin{equation}
<A_{k} (q)|A_{\mu } |0> = f_{A} m_{A} \varepsilon _{\mu }^{*} ,
\end{equation}
and $f_{A} $ is the decay constant of the emitted axial-vector meson $%
A_{k} $. The baryon-baryon matrix elements of the weak currents are defined as

\begin{equation} \label{10}
<B_{f} (p_{f} )|V_{\mu } |B_{i} (p_{i} )>\, =\bar{u}_{f} (p_{f} )[f_{1}
\gamma _{\mu } -\frac{f_{2} }{m_{i} } i\sigma _{\mu \nu } q^{\nu } +\frac{%
f_{3} }{m_{i} } q_{\mu } ]u_{i} (p_{i} ),
\end{equation}
and

\begin{equation}  \label{11}
<B_{f} (p_{f} )|A_{\mu } |B(p_{i} )>\, =\bar{u}_{f} (p_{f} )[g_{1} \gamma
_{\mu } \gamma _{5} -\frac{g_{2} }{m_{i} } i\sigma _{\mu \nu } q^{\nu }
\gamma _{5} +\frac{g_{3} }{m_{i} } q_{\mu } \gamma _{5} ]u_{i} (p_{i} ),
\end{equation}
where, $f_i$ and $g_i$ denote the vectot and axial-vector form factors as functions of $q^2$ \cite{35}. The factorizable amplitudes are thus given by

\begin{equation*}
A_{1}^{fac} =-\frac{G_{F} }{\sqrt{2} } F_{C} f_{A} c_{k} m_{A} [g_{1}^{B_{i}
,B_{f} } (m_{A}^{2} )-g_{2}^{B_{i} ,B_{f} } (m_{A}^{2} )\frac{m_{i} -m_{f} }{%
m_{i} } ],
\end{equation*}

\begin{equation*}
A_{2}^{fac} =\frac{G_{F} }{\sqrt{2} } F_{C} f_{A} c_{k} m_{A} [2g_{2}^{B_{i}
,B_{f} } (m_{A}^{2} )/m_{i} ],
\end{equation*}

\begin{equation*}
B_{1}^{fac} =\frac{G_{F} }{\sqrt{2} } F_{C} f_{A} c_{k} m_{A} [f_{1}^{B_{i}
,B_{f} } (m_{A}^{2} )+f_{2}^{B_{i} ,B_{f} } (m_{A}^{2} )\frac{m_{i} +m_{f} }{%
m_{i} } ],
\end{equation*}

\begin{equation}
B_{2}^{fac} =-\frac{G_{F} }{\sqrt{2} } F_{C} f_{A} c_{k} m_{A}
[2f_{2}^{B_{i} ,B_{f} } (m_{A}^{2} )/m_{i} ],
\end{equation}
where $F_C$ contains appropriate CKM factors and Clebsch-Gordan (CG)
coefficients and $c_k$ are QCD coefficients.

The baryon-baryon transition form factors $f_{i} $ and $g_{i} $ are evaluated in literature using the non-relativistic quark model (NRQM) \cite{33}  and the heavy quark effective theory (HQET) \cite{47}. The NRQM based form factors are calculated in Breit frame and include correction like the $q^2$ dependence of the form factors, the hard-gluon QCD contributions, and the wave-function mismatch. Later, in light of the fact that the form factors for heavy baryon-baryon transitions should also include constraints from the heavy quark symmetry, $1/m_{Q}$ correction to the form factors was introduced using HQET. It may be noted that the $\omc$ decays involving factorizable amplitudes only include $\omc^0 \to \x^0$ and $\omc^0 \to \x^-$ form factors which come out to be equal, numerically. The evaluated form factors using NRQM \cite{33} are given by
\ben \omc \to \x :~~~ f_1(0)=-0.23,~ f_2(0)=0.21 ,~ g_1(0)=0.14, ~g_2(0)=-0.019.
\en
Similarly, form factors calculation in HQET \cite{47} yields :
\ben \omc \to \x :~~~ f_1(0)=-0.34,~ f_2(0)=0.35 ,~ g_1(0)=0.10, ~g_2(0)=-0.020.
\en
We wish to remark here that numerical calculations of the factorizable  branching ratios we use dipole $q^2$ dependence following HQET constraints.

The decay constants of axial-vector mesons \cite{73,74,75,76,77} used for numerical evaluations are given by
\begin{center}
$f_{a_{1} }\approx f_{f_1}=0.221 $ GeV; $~~f_{K_{1}} (1270) =0.175$ GeV;
\end{center}
while decay constant for $\underline{K}_{1} (1.400) $ may be calculated by using relation $f_{\underline{K}_{1}} (1.400)/f_{K_{1}} (1.270) =\cot \theta _{1} $ \textit{i.e.}
\begin{center}
$f_{\underline{K}_{1}} (1.400) =-0.099 $ GeV, for $\theta _{1} =-58^{\circ} $; $f_{\underline{K}_{1}} (1.400) =-0.225$ GeV, for $\theta _{1} =-37^{\circ} $.
\end{center}

 It may also be noted that decay constants of axial-vector mesons are not so trivial to understand. As these may be effected by factors like, \textit{C}-parity/\textit{G}-parity conservations, mixing scheme and SU(3) breaking etc. For more details readers are referred to \cite{73,74,75,76,77}.

\subsection{Weak Transitions}

The flavor symmetric weak Hamiltonian \cite{40,45} for quark level process $q_{i}  +q_{j} \to q_{l}  +q_{m} $ can be expressed as,
\begin{equation}
H_{W} \cong V_{il} V_{jm}^* c_{-}(m_c)[\bar{B}^{[i,j]k} B_{[l,m]k} H_{[i,j]}^{[l,m]} ],
\end{equation}
where where $c_{-} = c_1 + c_2$ and the brackets, [~,~], represent the anti-symmetrization among the indices. The spurion transforms like $H_{[2,4]}^{[1,3]} $. We obtain the weak baryon-baryon matrix elements $a_{ij}$ for CKM favored and CKM suppressed modes from the following contraction:
\begin{equation}
H_{W} \cong a_W[\bar{B}^{[i,j]k} B_{[l,m]k} H_{[i,j]}^{[l,m]} ],
\end{equation}
where $a_W$ is weak amplitude. It is worth remarking here that the enhancement due to hard
gluon exchanges, coming through $c_{-}$, will effect weak baryon-baryon matrix element. Also, we ignore the long-distance QCD effects  reflected in the bound-state wave functions.

\subsection{ Axial-vector meson-baryon couplings}

In SU(4), Hamiltonian representing the strong transitions is given by
\ben H_{strong} =&\sqrt{2}& (g_D + g_{F}) (\frac{1}{2} \bar{B}^{[a,b]d} B_{[a,b]c} A_{d}^{c}) \non \\
&+&\sqrt{2} (g_{D}-g_F) ( \bar{B}^{[a,b]d} B_{[a,b]c} A_{d}^{c} ),
\en
where $B_{[a,b]c}  $,$ \bar{B}^{[a,b]d} $and $A_{d}^{c} $ are the baryon, anti-baryon, and axial-vector meson tensors, respectively and $g_{D} (g_{F} )$ are conventional D-(F)-type strong coupling constants \cite{30,33,37}.

 Experimentally, there are no measurements available for the axial-vector-meson-baryon coupling constants  for charm sector. Since, it is difficult to determine $g_{aNN}$ directly, a reasonable estimate could be obtained by using Goldberger-Treiman (GT) relation:
\ben
g_{NN\pi}=\frac{g_A m_N }{f_{\pi}},
\en
which relates pion-nucleon coupling $g_{NN\pi}$ with axial-vector coupling $g_A$ \cite{78, 79}. The GT relation exhibits direct relation between spontaneous chiral symmetry breaking and PCAC hypothesis at SU(2)$_L\times$SU(2)$_R$ or, with a possible extension, SU(3)$_L\times$SU(3)$_R$. Here, $g_A$ represents contribution to the dispersion relation of all the axial-vector states higher than the pion. In light of the PCAC, the heavier axial vector states contributing to $g_A$ must reproduce the pion pole at $q^2$=0. Thus, the combined contribution of all the heavier states may be replaced by an effective pole $a$ \textit{i.e} if we assume axial-vector dominance\footnote{As a consequence of spontaneous chiral symmetry breaking, Weinberg sum rules\cite{80} relate $m_{a_1}$ and $m_{\rho}$ by assuming vector and axial-vector dominance.} \cite{81,82} to get,
\ben g_{NNa_1}\approx \frac{g_A m_a^2 }{f_{a_1}}= 8.60
\en
for $g_A=1.26$ given by $\beta$ decay \cite{79}. To proceed forward, we use QCD sum rules analysis \cite{73} based
 \ben g_D=6.15~~ \mbox{and}~~  g_F=2.45 ~~\text{for} ~~\frac{g_D}{g_F}=2.5, \en
which in turn yields axial-vector meson-baryon strong coupling constants based on SU(4) symmetry.

We wish to pointed out that the SU(4) symmetry is badly broken, hence it would not be wise to use SU(4) symmetry based strong coupling constants for charm baryon decays. Therefore, we consider the SU(4) breaking effects in strong coupling constants by using the Coleman-Glashow null result \cite{83}. The tadpole mechanism can generate breaking effects, namely, the medium strong, the electromagnetic  and the weak effects, that transforms like an SU(3) octet, via a single symmetry breaking term. Thus, except for the tadpole term the hadronic Hamiltonian remains SU(3) invariant. In SU(3) octet, the strangeness changing scalar tadple $S_6$, transforming as sixth component of symmetry breaking octet, can be rotated away by unitary transformation. These strangeness changing effects produced by $S_6$ tadpole must vanish, leaving behind the electromagnetic and the weak effects. Khanna and Verma \cite{71} exploited the null result to obtain SU(3) broken baryon-baryon-pseudoscalar couplings. After validating SU(3) case they extended their results to SU(4), where symmetry breaking effects belong to similar regular representation \textbf{15}. In SU(4), the weak interaction Hamiltonian responsible for hadronic weak decays of charm baryons belongs to representation \textbf{20$^{''}$}. The tadpole term of the weak Hamiltonian belongs to representation \textbf{15}. In this case the charm changing effects generated through $S_9$ tadpole must vanish (for details see \cite{71}). We wish to remark here that the tadpole-type symmetry breaking effects does not include any additional parameter. The symmetry broken (SB) baryon-meson strong couplings are calculated by
\ben
g^{BB^{'}}_{A}\mbox{(SB)} = \frac{M_B +M_B^{'}}{2M_N} \frac{1}{\alpha_P} g^{BB^{'}}_{A}\mbox{(Sym)},
\en
where $g^{BB^{'}}_{A}$(Sym) is SU(4) symmetric couplings and the ratio of mass breaking terms $\alpha_P$ is \newpage
given by
 $$\alpha_P=\frac{\delta M_c}{\delta M_s}=\sqrt{\frac{3}{8}}\frac{m_c-m_u}{m_s-m_u}$$   \cite{71}. The obtained values of SB strong axial-vector meson-baryon coupling constants relevant for our calculation have been given in table \ref{t1}. For heavy baryon decays, it has been observed \cite{84} that mass independent strong couplings lead to smaller pole contributions. It is quite obvious that symmetry breaking will result in larger values for strong couplings as compared to symmetric ones due to mass dependence. Consequently, higher pole contributions would be expected. We also give the expressions for the $g^{BB^{'}}_{A}$ in terms of $g_D$ and $g_F$. However, we have given the absolute numerical values for the strong couplings where the actual sign would depend upon the conventions used and could be determined from their expressions in present case.

\subsection{Baryon Matrix Element}
In general, numerical evaluation of W-exchange terms (pole terms) involves weak matrix element of the form $\langle B_{f}\vert H_{W}^{PC} \vert B_{i} \rangle $. Riazuddin and Fayyazuddin \cite{72} has calculated this matrix element for noncharmed hyperon decays in the non-relativistic limit. Following their analysis one can obtain the matrix element for the charm baryons as a \textit{first approximation}. Though $\Omega_c$ is heavy and, therefore, outgoing quarks may have large momenta,
we use the non-relativistic approximation to get the first estimates of the baryonic matrix elements. The matrix element for the W-exchange process ($c+d \to s+ u$) can be expressed as
\ben
\mathcal{M}\approx \frac{G_F}{\sqrt{2}}V_{du}V_{cs}\big[\bar\psi_u(p_i^{'})\gamma_{\mu}(1-\gamma_5)\gamma_i^-\psi_d(p_i)\bar\psi_s(p_j^{'})\gamma_{\mu}(1-\gamma_5)\alpha_i^+\psi_c(p_j)+i \leftrightarrow j \big]
\en
 where $\psi$'s are Dirac spinors and $q=p_i-p_i^{'}=p_j^{'}-p_j$. The operators $\alpha^{+}_i$ convert $d \to u$ and $\gamma^{-}_j$ convert $c \to s$. In the leading non-relativistic approximation, only  terms corresponding to $\gamma^0$ and $\gamma^i \gamma_5$ have non zero limits, which are then reduced to only parity conserving part of $\mathcal{M}$. Thus, in leading non-relativistic approximation we have
 \ben \mathcal{M}^{PC}= \frac{G_F}{\sqrt{2}}V_{du} V_{cs} \sum_{i > j}(\gamma^{-}_i\alpha^{+}_j+\alpha^{+}_i \gamma^{-}_j ) (1-\sigma_i\cdot \sigma_j),\en
where \textbf{S$_i=$}$\mathbf{\sigma}_i/2$ are Pauli spinors representing spin of $i$th quark. Fourier transformation of the above expression gives the parity conserving weak Hamiltonian
\ben H^{PC}_W= \frac{G_F}{\sqrt{2}}V_{du} V_{cs} \sum_{i\neq j}\alpha^{+}_i \gamma^{-}_j (1-\sigma_i\cdot \sigma_j)\delta^3(r),\en
 following which, we can get a reasonable estimate of these terms. One can fix the scale by assuming the baryon overlap wave function to be flavor independent such that
\ben \label{30} \la \psi_f|\delta^3(r)|\psi_i\ra_{c} \approx \la \psi_f|\delta^3(r)|\psi_i\ra_{s},
\en
where $ \la \psi_f|\delta^3(r)|\psi_i\ra $ is baryon wave function overlap for corresponding flavor. We wish to remark here that (\ref{30}) leads to a well known SU(4) based relation that connects nonleptonic charmed baryon decays with hyperon decays. Since SU(4) is badly broken, a large mismatch between charm and strange baryon wave function overlaps would need a correction factor that has been practiced in many models based on different arguments \cite{39,44}.

\section{NUMERICAL RESULTS AND DISCUSSIONS}

 Summing over all the ingredients, the factorization and the pole contributions to different PV and PC amplitudes has been calculated. The numerical values of the possible factorizable contributions to weak decay amplitudes of $\Omega_c$ baryon in CKM-favored, suppressed and doubly-suppressed modes are shown in table \ref{t2}. Using the symmetry broken axial-vector meson-baryon couplings, we obtain the flavor independent pole amplitudes for all $\Omega_c$ baryon decays in CKM-favored, suppressed and doubly-suppressed modes as shown in column 2 of table \ref{t3}. We wish to emphasize that we have used only ground state $\frac{1}{2}^+$ intermediate baryon pole terms to estimate pole contributions. Adding factorizable and pole contributions, the branching ratios (BRs) and asymmetry parameters for the flavor independent case are predicted as shown in columns 3 and 5 of table \ref{t4}. Since, factorization contributes to only six of the $\omc$ decay modes, the remaining decay modes acquire contributions from pole amplitudes only. These branching ratios are given in column 2 of table \ref{t5}. We wish to point out that we use $\theta_{K_1} = -37^{\circ}$ as reference mixing angle, however, we also give results on mixing angle $-58^{\circ}$ for comparison. We summarize our results as follows:

\begin{enumerate}
\item The branching ratios of all the decay channels range from $10^{-3}$ to $10^{-7}$. The branching ratios of dominant modes are  $\mathcal{O} (10^{-3})$ $ \sim \mathcal{O}( 10^{-4})$.
\item Most of observed $\omc$ decay channels come from W-exchange processes, however, factorization processes contribute to only six of the decay channels.
\item The factorization contributions obtained from non-relativistic quark model (NRQM) and heavy quark effective theory (HQET) compare well without much discrepancies.
\item Among the $\omc$ decays acquiring contributions from both factorizable and pole amplitudes, the only possible CKM favored $(\Delta C = \Delta S = -1)$ decay mode $\oc \to \x^0 \kb^0$ has largest branching ratio $1.15 \t 10^{-3}$ for $\theta_{K_1}=-37^{\circ}$. The branching ratio increases further to a value $1.56 \t 10^{-3}$ for $\theta_{K_1}=-58^{\circ}$.
\item For $\oc \to \x^0 \kb^0$ decay, we find that the dominant contribution comes from factorizable amplitudes with pole contributions as low as $\sim 20-25\%$. It may be noted that color-suppressed factorizable amplitude interfere constructively with pole amplitude resulting in large branching ratio.

\item In CKM suppressed $(\Delta C = -1,~ \Delta S = 0)$ mode, the most dominant decay has Br($\oc \to \x^- a_1^+$)$\sim 1.00 \t 10^{-3}$ in HQET (though all the decays in this mode occur at comparable footing). We wish to point out that despite the CKM suppression and destructive interference betwen pole and factorization contributions, the  $\oc \to \x^- a_1^-$ decay is overly compensated by QCD enhancement ($c_1$).

\item The next order dominant decays in CKM suppressed mode are $\oc \to \x^0 a_1^0/\x^0 f_1$ with roughly comparable branching ratios. Here also, factorization and pole terms interfere constructively and destructively for the decays involving $f_1$ and $a_1$, respectively, though both are suppressed due to color suppression and small CG coefficients. It may be noted that $\oc \to \x^- a_1^+/\x^0 a_1^0/\x^0 f_1$ decays have dominant pole term contributions as compared to factorizable term contributions $\sim (20-40\%$).

\item As expected, the decay channels in Cabibbo doubly suppressed $(\Delta C = \Delta S = -1)$ modes have relatively smaller BRs of $\mathcal{O}(10^{-6}-10^{-7})$. We observe increase in branching ratio of the color suppressed $\omc \to \x^0 K_1^0$ decay despite of expected destructive interference between pole and factorization terms. We wish to remark here that the change of angle $\theta_{K_1}$ to $-58^{\circ}$ leads to smaller BR for $\omc \to \x^0 K_1^0$ though it increases for HQET case. The relative (signs) strengths of the S, P and D wave amplitudes may be attributed for the observed behavior. Similarly, We observe increase in branching ratio of the color favored mode $\omc \to \x^- K_1^+$ where factorization term appear to be dominant. It may also be noted that P wave amplitudes acquire larger magnitude in both these decays.

\item For $\omc$ decays acquiring contributions from pole terms (W-exchange diagrams) only, the CKM suppressed mode has BRs of $\mathcal{O}(10^{-4}-10^{-5})$. The dominant modes are $\omc \to \lb \kb^0/\lb \kbd^0$ with BRs of $\mathcal{O}(10^{-4})$. It may be noted that among decays arising from pole diagrams only, $\omc \to \lb \kb^0/\lb \kbd^0$ decays acquire most dominant pole amplitude contributions.

\item In case of CKM suppressed modes coming via pole diagrams only, the BRs are comparable to CKM suppressed modes of same category. The dominant decays are  $\omc \to p \kbd^-/\lb f_1^{'}$ with BR $\sim 1.00 \t 10^{-4}$. Branching ratio of all the remaining decays are of $\mathcal{O}(10^{-5})$. Despite of CKM suppression, BRs of these decays compete well with CKM suppressed modes due to higher pole contributions.

\item The absence of weak PV transition amplitudes ($b_{ij} $'s) lead to zero decay asymmetries for the decay modes coming from W-exchange processes only.

\item Also, we observe that mass dependence of strong coupling coming through SB effects result in larger strong couplings and hence, higher BRs.
\item Overall trend shows that he BRs of the decay modes involving $^3 P_1$ ($^1 P_1$) axial-vector states tend to increase (decrease) for $\theta_{K_1}=-58^{\circ}$.

\item All the decays involving non-strange meson in the final state have zero $u-$pole contributions except for $\oc \to \lb f_1^{'}$ decay which acquire contributions from both $u-$ and $s-$ channels. The highly suppressed decay modes $\oc \to \x^0 K_1^0/\x^- K_1^+$ have only $s$-pole contributions.
\item The decay modes consisting $b_1/h_1/h_1^{'}$ mesons in the final states are forbidden in  isospin limit.
\end{enumerate}

% \subsection{Flavor Dependent Effects}

In literatures, several attempts have been made to establish the fact that lifetimes of semileptonic and nonleptonic decays of heavy flavor baryons show a strong dependence on square of the baryon wave function overlap at the origin,  $\|\psi (0)\|^{2} $ \cite{43, 85,86,87}. In order to lower the discrepancy in theory and experiment, one needs to take in to account the variation of  $\|\psi (0)\|^{2} $ (being a dimensional quantity). For example, in case nonleptonic decays, inclusion of flavor dependence of hadron wave function at the origin  has resulted in good agreement between theory and experiment \cite{43, 88}. Following the analysis given in \cite{70}, we consider variation of $\|\psi (0)\|^{2} $ with flavor. It has been long advocated that a reliable estimate of wave function at origin of the ground state baryon can be obtained by experimental hyperfine splitting \cite{89}. A straightforward hyperfine splitting calculation, using constituent quark model, between $\Sigma_c$ and $\Lambda_c$ reveals
\begin{equation}
m_{\Sigma _{c} }-m_{\Lambda _{c} }=\frac{16 \pi}{9} \alpha _{s} (m_{c} ) \frac {(m_{c} -m_{u})}{m_c m_u^2} \vert \psi(0)\vert_{c}^{2} ,
\end{equation}
where we assume $\|\psi(0)\|^2_{\Sigma_c}=\|\psi(0)\|^2_{\Lambda_c}$. We obtain the flavor enhancement scale in strange and charm sectors from the following expression:
\begin{equation}
\frac{m_{\Sigma _{c}} -m_{\Lambda _{c}}}{m_{\Sigma} -m_{\Lambda} } =\frac{\alpha _{s} (m_{c} )}{\alpha _{s} (m_{s} )} \frac{m_{s} (m_{c} -m_{u} )|\psi (0)|_{c}^{2} }{m_{c} (m_{s} -m_{u} )|\psi (0)|_{s}^{2} } ,
\end{equation}
which yields
\begin{equation}
r\equiv \frac{|\psi (0)|_{c}^{2} }{|\psi (0)|_{s}^{2} } \approx 2.1,
\end{equation}
for the choice $\alpha _{s} (m_{c} )/\alpha _{s} (m_{s} )\approx 0.53$ \cite{42, 63}. Finally, we discuss the effects of this scale enhancement due to variation of spatial baryon wave function overlap on  branching ratios. The flavor dependent BRs for CKM-favored, suppressed and doubly-suppressed modes are evaluated using $|\psi (0)|^{2} $ variation. The numerical values of pole amplitudes only are given in column 3 of table \ref{t3}.  Consequently, the obtained numerical results for branching ratios and asymmetry parameters involving both factorizable and pole contributions are given in columns 4 and 6 of table \ref{t4}. Whereas the branching predictions for the processes involving pole contributions only are shown in column 3 of table \ref{t5}.   We wish to point out that the implications of variation of spatial baryon wave function overlap lead to flavor enhancement scale ratio ($r$) to $\sim$ 2. This may also be seen simply as a variation in $r$ from 1 to 2 for flavor dependent $|\psi(0)|^2$ owing to dimensionality arguments. It may also be noted that factorization hypothesis do not involve flavor dependent effects. In the absence of any experimental and theoretical information we compare our results with flavor independent BRs. We observe the following:

\begin{enumerate}
\item The variation of $|\psi(0)|^2$ has enhanced BRs of most of the decays roughly by a factor of four as compared to flavor independent BRs. Consequently, number of decay modes with BRs of $\mathcal{O} (10^{-3})$ $ \sim \mathcal{O}( 10^{-4})$ have become large.
\item Since, the factorization amplitudes remain unaffected by flavor dependent effects, the change in BRs in all the cases may be attributed due to flavor dependence effects on pole contributions.

\item In case of the $\omc$ decays involving both factorizable and pole amplitudes, the dominant decay channels $\oc \to \x^0 \kb^0 $ (for CKM favored) and $\oc \to \x^- a_1^+/\x^0 a_1^0/\x^0 f_1$ (for CKM suppressed mode) have BRs of $\mathcal{O} (10^{-3})$, which make them viable candidates for the experimental search. The highest Br$(\oc \to \x^- a_1^+)= 5.37 \t 10^{-3} $, where color enhancement has overcome CKM suppression. However, the branching ratio of color suppressed $\oc \to \x^0 \kb^0 $ decay comes out to be smaller \textit{i.e.} $2.41 \t 10^{-3}$. It is worth remarking that in spite of constructive interference between factorizable and pole amplitudes, the BR of $\oc \to \x^0 \kb^0 $ decay tend to be small in comparison to CKM suppressed mode. The reason being that the magnitude of the pole amplitude for CKM favored mode is smaller by an order when compared to CKM suppressed modes. However, the pole contributions to $\oc \to \x^0 \kb^0 $ decay arise up to $40-50\%$ because of flavor dependence which may further increase to $3.78 \t 10^{-3}$ for $\theta_{K_1}=-58^{\circ}$.

\item Unlike flavor independent case, CKM doubly suppressed decay modes $\omc \to \x^0 K_1^0/\x^- K_1^+$ show little change in BRs when flavor dependent effects are included. Comparable factorizable and pole terms add to the ambiguity of these decay modes. The relative magnitude and signs  of the S, P and D wave amplitudes become more important as it may be seen from variation in asymmetry parameter (both in sign and magnitude). Only experimental observation of these modes can provide a clear picture.

\item Among the $\omc$ decay modes arising through pole contributions only, the dominant decay modes with BRs of $\mathcal{O} (10^{-3})$ are $\oc \to \lb \kb^0/\lb \kbd^0$ with higher Br($\oc \to \lb \kbd^0$)= $2.13 \t 10^{-3}$. The BRs of all the remaining decay channels in CKM suppressed decay mode are enhanced to $\mathcal{O} (10^{-4})$.  However, the BRs may further increase or decrease with $\theta_{K_1}=-58^{\circ}$ for corresponding $K_1$ and $\kb$ modes, respectively. The comparable BRs of $\oc \to \lb \kb^0/\lb \kbd^0$ to that of CKM favored mode can be explained by dominant pole contributions to the former.

\item The flavor enhanced pole amplitudes has placed CKM doubly suppressed modes well in completion with CKM suppressed modes. The BRs of all these decays, namely, $\oc \to p \kbd^-/ n \kbd^0/\lb f_1^{'}/p \kb^-/ n \kb^0/\si^+a_1^-/\si ^0 a_1^0/\si ^- a_1^+$ have increased by an order of magnitude \textit{i.e.} to $ \mathcal{O} (10^{-4})$.  All these decay channels posses experimentally observable decay widths.
\end{enumerate}

\section{SUMMARY}

We have analyzed axial-vector meson emitting exclusive two-body nonleptonic weak decays of $\oc$ baryon for CKM-favored and suppressed modes in factorization and pole model approach. We have obtained the factorizable contributions by using the non-relativistic quark model (NRQM) \cite{33} and heavy quark effective theory (HQET)\cite{47} to evaluate the form factors $f_i$  and $g_i$. We expected that W-exchange diagrams could dominate $\oc$ weak decays and these are evaluated using pole model. The relevant baryon matrix elements of the weak Hamiltonian have been calculated which determine the pole term with short distance QCD corrections. Also, we have observed that mass dependence (SB effects) of strong couplings turns out to be crucial in deciding pole contributions to heavy baryon decays. These effects can be important specifically in the decays coming from the W-exchange process (pole diagram) only. Non-relativistic evaluation of weak matrix element involving PC weak Hamiltonian has been carried out for flavor independent and flavor dependent cases. We have predict BRs of $\oc$ decays for the cases  a) involving both factorization and pole amplitudes and b) arising via pole amplitudes (W-exchange diagram) only. We list our results as follows:
\begin{enumerate}

 \item For the flavor independent case, the only dominant decay mode $\oc \to \x^- a_1^+$ has branching ratio of $\mathcal{O}(10^{-3})$. The next order dominant modes are  $\oc \to \x^0 \kb^0/\x^0 a_1^0/\x^0 f_1$. All these decay modes consist interference of pole and factorizable contributions. In $\oc \to \x^- a_1^+$ decay, dominant contribution comes from factorization term while in rest of the decay channels pole contributions dominate. For the decay arising from pole amplitudes only, the $\oc \to \lb \kb^0/\lb f_1^{'}$ has branching ratios of $\mathcal{O}(10^{-4})$.
 \item For the flavor dependent case, we consider variation of spatial baryon wave function overlap at the origin \textit{i.e.} $|\psi(0)|^2$ with flavor. We observe that the introduction of flavor dependence has raised the BRs of all the decays roughly by a factor of four. A number of dominant modes  $\oc \to \x^- a_1^+/\x^0 \kb^0/\x^0 a_1^0/\x^0 f_1/\lb \kb^0/\lb \kbd^0$ now have BRs of $\mathcal{O}(10^{-3})$. All these decay channel fall in the limit of experimental reach.
 \item We wish to remark here that most of the decay channels in $\oc$ decay only through the W-exchange diagram; moreover, the W-exchange contributions dominate in rest of the process, with some exception. Observation of such decays would shed some light on mechanism of W-exchange effects in these decay modes.
\end{enumerate}
A conventional concept expects the $p$-wave emitting decays to be kinematically suppressed; however, we find that BRs of axial-vector emitting decays of $\oc$ are comparable to the experimentally observed two-body $s$-wave meson emitting decays of charm baryons. We hope this would generate ample interest in experimental search of these decay modes.

\acknowledgments
One of the authors (RD) would like to thank Prof. R.C. Verma for helpful discussions. This research is funded by the National Research Foundation of Korea (NRF)
grant funded by Korea government of the Ministry of Education, Science and
Technology (MEST) (No. 2011-0017430) and (No. 2011-0020333).

%\newpage

\begin{table}[h]
\captionof{table} {Expressions of strong-coupling constants [ SB = Sym. $\times( \frac{M_B +M_B^{'}}{2M_N \alpha_P}) $] and their absolute numerical values at $\theta_{K_1}=-37^{\circ} (-58 ^{\circ})$.}
\label{t1}
\renewcommand{\arraystretch}{1.15}
\begin{tabular}{|c|c|c|}\hline
\multicolumn{2}{|c|}{Strong Couplings } & Absolute Values   \\
\multicolumn{2}{|c|}{$g^{BB^{'}}_{A}$ $\times( \frac{M_B +M_B^{'}}{2M_N \alpha_P} $) } & $|g^{BB^{'}}_{A}$(SB)$|$ \\ \hline
$g^{\lb p}_{K_1}$ &$(\sqrt{3} g_D + \frac{g_F}{\sqrt{3}})\sin \theta_{K_1}$& 5.13 (7.23) \\
$g^{\si^0 p}_{K_1}$& $(-g_D+g_F)\sin \theta_{K_1}$ &2.53 (3.56)\\
$g^{\lb p}_{\k1d}$&$(\sqrt{3} g_D + \frac{g_F}{\sqrt{3}})\cos \theta_{K_1}$ &6.81 (4.52) \\
$g^{\si^0 p}_{\k1d}$& $(-g_D+g_F)\cos \theta_{K_1}$ &3.35 (2.23) \\
$g^{\lb n}_{K_1}$&$-(\sqrt{3} g_D + \frac{g_F}{\sqrt{3}})\sin \theta_{K_1}$&5.14 (7.24) \\
$g^{\si^0 n}_{K_1}$& $(-g_D+g_F)\sin \theta_{K_1}$&2.53 (3.57) \\
$g^{\lb n}_{\k1d}$&$-(\sqrt{3} g_D + \frac{g_F}{\sqrt{3}})\cos \theta_{K_1}$&6.82 (4.52) \\
$g^{\si^0 n}_{\k1d}$& $(-g_D+g_F)\cos \theta_{K_1}$&3.58 (2.23) \\
$g^{\x^0 \lb}_{K_1}$ &$(-\sqrt{3}g_D+\frac{g_F}{\sqrt{3}}) \sin \theta_{K_1} $ &0.54 (0.76) \\
$g^{\x^- \lb}_{K_1}$ &$(\sqrt{3}g_D-\frac{g_F}{\sqrt{3}}) \sin \theta_{K_1} $ & 0.54 (0.76) \\
$g^{\x^0 \lb}_{\k1d}$&$(-\sqrt{3}g_D+\frac{g_F}{\sqrt{3}}) \cos \theta_{K_1} $  &0.72 (0.47)\\
$g^{\si^0 \x^{0(-)}}_{K_1}$ &$-(g_D+g_F) \sin \theta_{K_1}$&6.92 (9.75) \\
$g^{\si^0 \x^{0(-)}}_{\k1d}$ &$-(g_D+g_F) \cos \theta_{K_1}$&9.18 (6.10) \\
$g^{\si^+ \x^0 }_{K_1}$& $-\sqrt{2}(g_D+g_F)\sin \theta_{K_1}$&9.77 (13.76) \\
$g^{\si^+ \x^0 }_{\k1d}$& $-\sqrt{2}(g_D+g_F)\cos \theta_{K_1}$&12.96 (8.60) \\
$g^{\omc \xic}_{K_1}$ & $ \frac{-2g_F}{\sqrt{3}} \sin \theta_{K_1}$&11.77 (16.58) \\
$g^{\omc \xicd}_{K_1}$& $ 2 g_D \sin \theta_{K_1}$&8.27 (11.72) \\
 $g^{\omc \xic}_{\k1d}$ &$ \frac{-2g_F}{\sqrt{3}} \cos \theta_{K_1}$&~15.62 (10.34)\\
$g^{\omc \xicd}_{\k1d}$&$ 2 g_D \cos \theta_{K_1}$ &11.00 (7.30) \\
$g^{\lb \lb}_{f_1}$& $2(g_D-\frac{2g_F}{3})$&3.92 \\
 $g^{\lb \lb}_{f_1^{'}}$&$-\sqrt{2}(g_D+\frac{g_F}{3})$&~7.57 \\
 $g^{\si^0 \lb}_{f_1/f_1^{'}}$&0&~0\\
$g^{\lb \si^+}_{a_1}$ &$\frac{2g_F}{\sqrt{3}}$&8.72 \\
$g^{\lb \si^{0(-)}}_{a_1}$ &$-\frac{2g_F}{\sqrt{3}}$&8.74 \\
$g^{\si^0 \si^0}_{a_1}$ &0&0 \\
$g^{\si^0 \si^{+(-)}}_{a_1}$& $2g_D$&6.22 \\
$g^{\x^0 \x^0}_{{a_1}/{f_1}}$ & $g_D-g_F$& ~5.18 \\
$g^{\x^0 \x^-}_{a_1^+}$& $\sqrt{2}(g_D-g_F)$&~7.35\\
$g^{\omc \omc}_{a_1/f_1}$ &0&0\\
$g^{\omc \omc}_{f_1^{'}}$ &$-2\sqrt{2}g_D$&~19.92\\ \hline
\end{tabular}
\end{table}

\begin{table}[h]
\renewcommand{\arraystretch}{1.15}
\captionof{table} {Factorizable amplitudes (in units of $\frac{G_{F}}{ \sqrt{2}}  V_{uq} V_{cq}^{*}$) to $\oc$ decays for CKM-favored, CKM-suppressed and CKM- doubly-suppressed modes.}
\label{t2}
\begin{tabular}{|c|c|c|c|c|c|}
\cline{1-6}
Deacys                                  & Model   & \multicolumn{4}{c|}{Factorizable amplitudes\footnote{The factorizable amplitudes are independent of mixing angle $\theta_{K_1}$ for the decays emitting $K_1(1270)$ meson. Since, decay constant of $K_1(1270)$ do not depend upon the $K_1(1270)- K_1(1400)$ mixing angle, which essentially affects the decay constant of $K_1(1400)$.}}                                                                                                      \\ \cline{3-6}
                                           & \cite{33}\cite{47} & \multicolumn{1}{c|}{$A_1^{fac}$} & \multicolumn{1}{c|}{$A_2^{fac}$} & \multicolumn{1}{c|}{$B_1^{fac}$} & \multicolumn{1}{c|}{$B_2^{fac}$} \\ \hline
\multicolumn{6}{|l|}{Cabbibo-favored $\Delta C = -1, \Delta S= 0$ mode}                                                                                                                        \\ \hline
\multicolumn{1}{|c|}{$\oc \to \x^0 \kb^0$} & NRQM  & 0.033                   & 0.0017                  & $-0.0027$             & 0.052                    \\
\multicolumn{1}{|c|}{}                     & HQET  & 0.025                     & 0.0034                   & $-0.0060$            & 0.087                     \\ \hline
\multicolumn{6}{|l|}{CKM-suppressed $\Delta C = -1, \Delta S= 0$ mode}                                                                                                                     \\ \hline
\multicolumn{1}{|c|}{$\oc \to \x^0 a_1^0$} & NRQM  & $-0.026$                         & $-0.0013$                        & $0.021$                          & $-0.039$                         \\
\multicolumn{1}{|c|}{}                     & HQET  & $-0.018$                         & $-0.0026$                        & $0.045$                          & $-0.065$                         \\ \hline
\multicolumn{1}{|c|}{$\oc \to \x^0 f_1$}   & NRQM  & $0.029$                          & $0.0015$                         & $-0.024$                         & $0.045$                          \\
\multicolumn{1}{|c|}{}                     & HQET  & $0.022$                          & $0.0029$                         & $-0.052$                         & $0.075$                          \\ \hline
\multicolumn{1}{|c|}{$\oc \to \x^- a_1^+$} & NRQM  & $-0.090$                         & $-0.0046$                        & $0.072$                          & $-0.14$                         \\
\multicolumn{1}{|c|}{}                     & HQET  & $-0.068$                          & $-0.0092$                        & $0.160$                          & $-0.23$                         \\ \hline
\multicolumn{6}{|l|}{CKM-doubly-suppressed $\Delta C =-\Delta S= -1$ mode}                                                                                                                 \\ \hline
\multicolumn{1}{|c|}{$\oc \to \x^0 K_1^0$} & NRQM  & $0.033$                   & $0.017$                 & $-0.027$              & $0.052$                 \\
\multicolumn{1}{|c|}{}                     & HQET  & $0.025$                 & $0.034$                 & $-0.062$               & $0.087$                 \\ \hline
\multicolumn{1}{|c|}{$\oc \to \x^- K_1^+$} & NRQM  & $-0.082$               & $-0.0042$             & $0.068$                & $-0.128$               \\
\multicolumn{1}{|c|}{}                     & HQET  & $-0.062$                 & $-0.0083$             & $0.149$                & $-0.214$                \\ \hline
\end{tabular}
\end{table}

\begin{table}[h]
\renewcommand{\arraystretch}{1.15}
\captionof{table} {Pole amplitudes (in units of $\frac{G_{F}}{ \sqrt{2}}  V_{uq} V_{cq}^{*}$) of all $\oc$ decays for CKM-favored, CKM-suppressed and CKM- doubly-suppressed  modes at $\theta_{K_1}=-37^{\circ} (-58 ^{\circ})$. Flavor dependent pole contributions include effects of $|\psi(0)|^2$ variation.}
\label{t3}
\begin{tabular}{|c|c|c|}
\hline Decays &\multicolumn{2}{c|}{Pole Amplitudes } \\ \cline{2-3}
 & Flavor independent  & Flavor dependent   \\ \hline
\multicolumn{3}{|l|}{CKM-favored ($\Delta C =  \Delta S = -1$) mode }\\ \hline
$\oc \to \x^0 \kb^0$   &$ -0.026$  ($-0.036$)&$-0.054$  ($-0.076$)\\ \hline
\multicolumn{3}{|l|}{CKM-suppressed ($\Delta C = -1, \Delta S = 0$) mode }\\ \hline
$\oc \to \x^0 a_1^0$ &$-0.20$  & $-0.42$ \\ \hline
 $\oc \to \x^0 f_1$ &$-0.20$  &$-0.42$  \\ \hline
 $\oc \to \x^- a_1^+$ & $-0.28$ & $-0.59$ \\ \hline
$\oc \to \lb \kb^0$ &$0.12$ ($0.17$) &$0.27$ ($0.36$) \\ \hline
$\oc \to \lb \kbd^{0}$ &$-0.16$ ($-0.11$)  &$-0.34$ ($-0.23$) \\ \hline
$\oc \to \si^+ K_1^-$ & $0.049$ ($0.069$) & $0.10$ ($0.14$)\\ \hline
$\oc \to \si^+ \k1d^-$  & $-0.065$ ($-0.043$) & $-0.14$ ($-0.090$) \\ \hline
 $\oc \to \si^0 \kb^0$ &$0.034$ ($0.048$) &$0.072$ ($0.10$) \\ \hline
$\oc \to \si^0 \kbd^{0}$  & $-0.046$ ($-0.030$)& $-0.096$ ($-0.064$)\\ \hline
\multicolumn{3}{|l|}{CKM-doubly-suppressed ($\Delta C = - \Delta S = -1$) mode }\\ \hline
 $\oc \to \x^0 K_1^0$ &$0.015$ ($0.021$) & $0.031$ ($0.044$) \\ \hline
$\oc \to \x^- K_1^+$ &$-0.015$ ($-0.021$) & $-0.031$ ($-0.044$) \\ \hline
$\oc \to p K_1^-$ & $0.22$ ($0.30$) & $0.45$  ($0.64$)  \\ \hline
$\oc \to p \k1d^{-}$ & $-0.29$ ($-0.19$) & $-0.60$ ($-0.40$) \\ \hline
$\oc \to n K_1^0$ &  $-0.22$ ($-0.30$) & $-0.45$  ($-0.64$)  \\ \hline
 $\oc \to n \k1d^{0}$& $0.29$ ($0.19$) & $0.60$ ($0.40$) \\ \hline

 $\oc \to \lb f_1$&$-0.11$  & $-0.22$ \\ \hline
  $\oc \to \lb f_1^{'}$&$0.34$  & $0.71$ \\ \hline
  $\oc \to \si ^+ a_1^-$& $0.24$ & $0.50$ \\ \hline
$\oc \to \si ^0 a_1^0$ & $-0.24$ & $-0.50$ \\ \hline
 $\oc \to \si ^- a_1^+$& $-0.24$ & $-0.50$ \\ \hline
\end{tabular}
\end{table}

\begin{table}[h]
\renewcommand{\arraystretch}{1.15}
\captionof{table} {Branching ratios and asymmetry parameters of $\oc$ decays acquiring contributions from both factorization and pole amplitudes for CKM-favored, CKM-suppressed and CKM-doubly-suppressed modes at $\theta_{K_1}=-37^{\circ} (-58 ^{\circ})$. Flavor dependent results include effects of $|\psi(0)|^2$ variation.}
\label{t4}
\begin{tabular}{|c|c|l|l|l|l|}
\cline{1-6}
Deacys                              &  Model     & \multicolumn{2}{c|}{Branching ratios}                                                   & \multicolumn{2}{c|}{Asymmetry '$\alpha$'}                                       \\ \cline{3-6}
                                           & \cite{33}\cite{47} & \multicolumn{1}{c|}{Flavor} & \multicolumn{1}{c|}{Flavor} & \multicolumn{1}{c|}{Flavor} & \multicolumn{1}{c|}{Flavor} \\
                                           &  & \multicolumn{1}{c|}{independent } & \multicolumn{1}{c|}{dependent} & \multicolumn{1}{c|}{independent} & \multicolumn{1}{c|}{dependent} \\ \hline
\multicolumn{6}{|l|}{CKM-favored $\Delta C = -1, \Delta S= 0$ mode}                                                                                                                                                \\ \hline
\multicolumn{1}{|c|}{$\oc \to \x^0 \kb^0$} & NRQM  & $1.15 \t 10^{-3}$                         ($1.56 \t 10^{-3}$)    & $2.42 \t 10^{-3}$                         ($3.78 \t 10^{-3}$)                      & $0.39$ ($0.34$)                    & 0.28 (0.22)                         \\
\multicolumn{1}{|c|}{}                     & HQET  & $0.98 \t 10^{-3}$                         ($1.34 \t 10^{-3}$)          & $2.11 \t 10^{-3}$                         ($3.37 \t 10^{-3}$)                       & $0.63$ ($0.55$)                   & 0.46 (0.36)                                                 \\ \hline
\multicolumn{6}{|l|}{CKM-suppressed $\Delta C = -1, \Delta S= 0$ mode}                                                                                                                                             \\ \hline
\multicolumn{1}{|c|}{$\oc \to \x^0 a_1^0$} & NRQM  & $5.90\t 10^{-4}$                                                        & $3.00 \t 10^{-3}$                                                     & $-0.13$                 & $-0.058$                              \\
\multicolumn{1}{|c|}{}                     & HQET  & $6.40 \t 10^{-4}$                                                        & $3.11 \t 10^{-3}$                                                     & $-0.20$                                  & $-0.091$                              \\ \hline
\multicolumn{1}{|c|}{$\oc \to \x^0 f_1$}   & NRQM  & $7.96 \t 10^{-4}$                                                         & $3.05 \t 10^{-3}$                                                      & $0.091$                                & $0.045$                               \\
\multicolumn{1}{|c|}{}                     & HQET  & $7.51 \t 10^{-4}$                                                         & $2.96 \t 10^{-3}$                                                      & $0.15$                                & $0.076$                               \\ \hline
\multicolumn{1}{|c|}{$\oc \to \x^- a_1^+$} & NRQM  & $7.58 \t 10^{-4}$                                                        & $4.88 \t 10^{-3}$                                                    & $-0.38$                                 & $-0.15$                              \\
\multicolumn{1}{|c|}{}                     & HQET  & $9.90\t 10^{-4}$                                                         & $5.37 \t 10^{-3}$                                                     & $-0.51$                                 & $-0.23$ \\ \hline
\multicolumn{6}{|l|}{CKM-doubly-suppressed $\Delta C =-\Delta S= -1$ mode}                                                                                                                                         \\ \hline
\multicolumn{1}{|c|}{$\oc \to \x^0 K_1^0$} & NRQM  & $5.15 \t 10^{-7}$ ($3.42 \t 10^{-7}$)
 & $2.41 \t 10^{-7}$ ($3.40 \t 10^{-7}$)                     & $0.74$ ($0.69$)                     & $0.16$ ($-0.70$)                     \\
\multicolumn{1}{|c|}{}                     & HQET  & $6.10 \t 10^{-7}$ ($5.16 \t 10^{-7}$)                      & $5.54 \t 10^{-7}$ ($8.15 \t 10^{-7}$)                     & $0.55$ ($0.23$)                     & $-0.38$ ($-0.78$)                     \\ \hline
\multicolumn{1}{|c|}{$\oc \to \x^- K_1^+$} & NRQM  & $3.40 \t 10^{-6}$ ($2.35 \t 10^{-6}$)                     &  $4.93 \t 10^{-6}$ ($4.13 \t 10^{-6}$)                                    & $0.66$ ($0.69$)                       & $0.72$ ($0.74$)                    \\
\multicolumn{1}{|c|}{}                     & HQET  & $4.70 \t 10^{-6}$ ($4.10 \t 10^{-6}$)                                           &  $3.67 \t 10^{-6}$ ($3.04 \t 10^{-6}$)                                     & $0.77$ ($0.73$)                       & $0.64$ ($0.41$)                    \\ \hline
\end{tabular}
\end{table}

\begin{table}[h]
\captionof{table} {Branching ratios of $\oc$ for CKM-suppressed and CKM-doubly-suppressed modes at $\theta_{K_1}=-37^{\circ} (-58 ^{\circ})$ acquiring contributions from pole amplitudes only. Flavor dependent branching ratios include effects of $|\psi(0)|^2$ variation.}
\label{t5}
\renewcommand{\arraystretch}{1.15}
\begin{tabular}{|c|l|l|}
\hline
Decays & Flavor independent BRs  & Flavor dependent BRs  \\ \hline
\multicolumn{3}{|l|}{CKM-suppressed ($\Delta C = -1, \Delta S= 0$) mode }\\ \hline
$\oc \to \lb \kb^0$ &$3.96\t 10^{-4}$ ($7.89\t 10^{-4}$) &$1.74\t 10^{-3}$ ($3.47 \t 10^{-3}$) \\ \hline
$\oc \to \lb \kbd^{0}$ &$4.83\t 10^{-4}$ ($2.12 \t 10^{-4}$)  &$2.13\t 10^{-3}$ ($9.37 \t 10^{-4}$) \\ \hline
$\oc \to \si^+ K_1^-$ & $5.50\t 10^{-5}$ ($1.09 \t 10^{-4}$) & $2.42\t 10^{-4}$ ($4.82 \t 10^{-4}$)\\ \hline
$\oc \to \si^+ \k1d^-$  & $5.94\t 10^{-5}$ ($2.62 \t 10^{-4}$) & $2.62\t 10^{-4}$ ($1.15 \t 10^{-4}$) \\ \hline
 $\oc \to \si^0 \kb^0$ &$2.72\t 10^{-5}$ ($5.41 \t 10^{-5}$) &$1.20\t 10^{-4}$ ($2.39 \t 10^{-4}$) \\ \hline
$\oc \to \si^0 \kbd^{0}$  & $2.92\t 10^{-5}$ ($1.28 \t 10^{-5}$)& $1.29\t 10^{-4}$ ($5.67 \t 10^{-5}$)\\ \hline
 \multicolumn{3}{|l|}{CKM-doubly-suppressed ($\Delta C = - \Delta S = -1$) mode }\\ \hline
$\oc \to p K_1^-$ & $7.75\t 10^{-5}$ ($1.54 \t 10^{-4}$) & $3.42\t 10^{-4}$  ($6.79 \t 10^{-4}$)  \\ \hline
$\oc \to p \k1d^{-}$ & $1.04\t 10^{-4}$ ($4.60\t 10^{-5}$) & $4.60\t 10^{-4}$ ($2.03 \t 10^{-4}$) \\ \hline
$\oc \to n K_1^0$ & $7.72\t 10^{-5}$ ($1.53\t 10^{-4}$) &$3.41\t 10^{-4}$ ($6.77\t 10^{-4}$) \\ \hline
 $\oc \to n \k1d^{0}$& $1.04\t 10^{-5}$ ($4.58\t 10^{-5}$) & $4.58\t 10^{-4}$ ($2.02\t 10^{-4}$) \\ \hline
  $\oc \to \lb f_1$&$1.55\t 10^{-5}$  & $6.83\t 10^{-5}$ \\ \hline
  $\oc \to \lb f_1^{'}$&$1.00\t 10^{-4}$  & $4.38\t 10^{-4}$ \\ \hline
  $\oc \to \si ^+ a_1^-$& $7.75\t 10^{-5}$ & $3.42\t 10^{-4}$ \\ \hline
$\oc \to \si ^0 a_1^0$ & $7.74\t 10^{-5}$ & $3.41 \t 10^{-4}$ \\ \hline
 $\oc \to \si ^- a_1^+$& $7.72\t 10^{-5}$ & $3.40 \t 10^{-4}$ \\ \hline
\end{tabular}
\end{table}

\end{document}